\documentstyle[12pt,aps,prl]{revtex}
\begin{document}

\draft 

\title{Dynamical heterogeneities in a supercooled Lennard-Jones
liquid}

\author{Walter Kob$^1$, Claudio Donati$^2$, Steven J. Plimpton$^3$, 
Peter H. Poole$^4$, and Sharon C.  Glotzer$^{2}$} 

\address{$^1$ Institut f\"ur Physik, Johannes Gutenberg-Universit\"at,
Staudinger Weg 7, D-55099 Mainz, Germany}

\address{$^2$ Center for Theoretical and Computational Materials
Science, and Polymers Division, National Institute of Standards and
Technology, Gaithersburg, Maryland, USA 20899}

\address{$^3$ Parallel Computational Sciences Department, 
Sandia National Laboratory, Albuquerque, NM 87185-1111}

\address{$^4$Department of Applied Mathematics,
University of Western Ontario, London, Ontario N6A~5B7, Canada}

\date{\today}

\maketitle

\begin{abstract}
We present the results of a large scale molecular dynamics computer
simulation study in which we investigate whether a supercooled
Lennard-Jones liquid exhibits dynamical heterogeneities. We evaluate
the non-Gaussian parameter for the self part of the van Hove
correlation function and use it to identify ``mobile'' particles. We
find that these particles form clusters whose size grows with
decreasing temperature.  We also find that the relaxation time of the
mobile particles is significantly shorter than that of the bulk, and
that this difference increases with decreasing temperature.

\end{abstract}


\narrowtext

Recent NMR experiments have shown that the relaxation in supercooled
liquids is not homogeneous, i.e. that there are regions in space in
which the relaxation of the particles is significantly faster (or
slower) than the average relaxation of the
system~\cite{schmidt91}. Subsequently, this result has been supported
by optical spectroscopy, forced Rayleigh scattering and further NMR
experiments~\cite{hetero_exp}.  However, these types of experiments
are unable to determine the nature of these ``dynamical
heterogeneities,'' and consequently details such as size are unknown.
Extrapolations of probe size sensitivity to heterogeneous dynamics
indicate a typical heterogeneity size on the order of 2 -- 5 nm in the
vicinity of $T_g$\cite{size-ediger}.

Dynamical heterogeneities have also been observed in computer
simulations~\cite{harrowell}. However, these simulations were
restricted to two dimensions and since it is expected that the dynamics
of particles in two and three dimensions is significantly different, it
is not clear whether the dynamical heterogeneities observed in these
simulations have a counterpart in three dimensions. By analyzing
the trajectories of monomers in a Monte Carlo simulation of a dense
($d=3$) polymer melt, Heuer and Okun~\cite{heuer97} showed that in this
system dynamical heterogeneities occur on short length scales, but the
nature of the heterogeneities was not explored in detail. Thus despite
the experimental evidence for the existence of dynamical
heterogeneities, their microscopic properties are unknown and thus
phenomenological models are often used to interpret experimental
results~\cite{cicerone97}.  In this Letter, we study a simple,
glass-forming liquid to investigate whether dynamical heterogeneities
can be observed in a 3-d system and, if so, to determine their properties.

We investigate a binary (80:20) mixture of 8000 Lennard-Jones particles
consisting of two species of particles, $A$ and $B$.  The interaction
between two particles of type $\alpha, \beta \in \{A,B\}$ is given by
$V_{\alpha\beta}(r)=4\epsilon_{\alpha\beta}
[(\sigma_{\alpha\beta}/r)^{12} -(\sigma_{\alpha\beta}/r)^6]$ with
$\epsilon_{AA}=1.0$, $\sigma_{AA}=1.0$, $\epsilon_{AB}=1.5$,
$\sigma_{AB}=0.8$, $\epsilon_{BB}=0.5$, and $\sigma_{BB}=0.88$, with a
cuttoff radius of $2.5\sigma_{\alpha\beta}$.  Note that the $AB$
interaction is stronger than both the $AA$ and $BB$ interactions, a
fact which will be important in the subsequent discussion of the
results.  We report all quantities in reduced units, i.e. length in
units of $\sigma_{AA}$, temperature $T$ in units of
$\epsilon_{AA}/k_B$, and time $t$ in units of
$\sqrt{\sigma_{AA}^2m/\epsilon_{AA}}$, where $m$ is the mass of either
an $A$ or $B$ particle.  We study the system at 10 different values of
$T$ ranging between 0.550 and 0.451. At each $T$, the system was
equilibrated for a time longer than the $\alpha$-relaxation time before
evaluating the quantities presented below. At the lowest $T$,
quantities were evaluated for over $4\times 10^6$ time steps.  All
simulations were carried out in the microcanonical ensemble.  More
details on the simulation can be found in Ref.~\cite{kob_plimpton}.

The dynamics of this model has been characterized in detail in
previous simulations performed at different temperatures and at constant
density~\cite{kob}.  In particular, it was found that at low $T$ the
dynamics is described well by mode-coupling theory~\cite{gotze92} with
a critical temperature $T_c\approx 0.435$ and a critical pressure $P_c
\approx 3.03$.  In the present work we approach the point $(T_c,P_c)$
via a different path than that used in Ref.~\cite{kob}, on a straight
line in the $T-P$ plane along which density increases with decreasing
$T$ \cite{path}.  It has been shown~\cite{kob_plimpton} that along
this path of approach to the critical point, the behavior of the
relaxation dynamics is very similar to that found along the
constant-density path of the previous simulation~\cite{kob}, providing
evidence that the thermodynamic path of approach to the critical point
does not significantly change the nature of the divergence of the
relaxation time, and hence how the system vitrifies.  Hence, we expect
the results presented here to be independent of the details of the
approach to the glass transition, and so in the following we use $T$
alone to characterize the different state points.

To detect the presence of dynamical heterogeneities, we investigate the
time dependence of the self part $G_s(r,t)$ of the van~Hove correlation
function~\cite{hansenmcdonald86} for the $A$ particles, where $r$ is
the distance traveled by a particle in a time $t$.  To a first
approximation $G_s(r,t)$ has a Gaussian form but deviations from this
form at intermediate times have been observed in simulations of glass
forming liquids~\cite{kob,thirumalai93,harrowell1} and are thought to
reflect the presence of dynamical heterogeneities~\cite{harrowell2}.
Such deviations can be characterized by the non-Gaussian parameter of
$G_s(r,t)$, $\alpha_2(t)=3\langle r^{4}(t) \rangle/5\langle r^{2}(t)
\rangle ^{2}-1$ \cite{rahman64}.  Fig.~\ref{fig1} shows the time
dependence of $\alpha_2$ for the $A$ particles at three different
temperatures. We find that: (i) on the time scale at which the motion
of the particles is ballistic, $\alpha_2$ is zero; (ii) upon entering
the time scale of the $\beta$-relaxation $\alpha_2$ starts to increase;
and (iii) on the time scale of the $\alpha$-relaxation, $\alpha_2$
decreases to its long time limit, zero.  We observe that the maximum
value of $\alpha_2$ increases with decreasing $T$, which is evidence
that the dynamics of the liquid becomes more heterogeneous with
decreasing $T$.  Furthermore, we find that the time $t^*$ at which this
maximum is attained also increases with decreasing $T$.

To determine the reason for the strong increase of $\alpha_2$ in the
$\beta$-relaxation regime, we compare $G_s(r,t)$ with the distribution
that is obtained from the Gaussian approximation, i.e. by assuming
that $G_s(r,t)$ is given by $G_s^{g} (r,t)=(3/2\pi \langle
r^2(t)\rangle)^{3/2}\exp(-3r^2/2\langle r^2(t) \rangle)$, where
$\langle r^2(t) \rangle$ is the mean squared displacement of the
particles.  In Fig.~\ref{fig2} we show $(G_s(r,t)-G_s^{g}(r,t))/
G_s^{g}(r,t)$ for $t=t^*$, where $t^*$ depends on $T$ (see
Fig.~\ref{fig1}). For small and intermediate values of $r$ ($r\leq
0.6$) the relative difference between $G_s^{g}$ and $G_s$ is less than
a factor of three. However, for larger $r$, $G_s^{g}$ underestimates
$G_s$ significantly.  The discrepancy increases strongly with
decreasing $T$ in that the normalized difference becomes as large as
$10^8$ at the lowest $T$ (see inset of Fig.~\ref{fig2}). Thus we find
that in the supercooled liquid there is a significant number of
particles that have moved farther than would be expected from the
Gaussian approximation~\cite{thirumalai93}.  We define $r^*$ as the
larger value of $r$ such that $G_s(r^*,t^*)= G_s^{g}(r^*,t^*)$, i.e.
$r^*$ is the value of $r$ at which the normalized difference starts to
become positive and very large (see Fig.~\ref{fig2}).  {\it We thereby
define ``mobile particles'' as $A$ particles that have moved farther
than a distance $r^*$ within a time $t^*$.}  With this definition, the
total number of mobile particles at any $T$ studied is a few hundred
(out of 6400 $A$ particles), and thus constitute approximately 5\% of
the system.  We note that we find the results presented below
concerning the properties of the mobile particles to be relatively
insensitive to the details of this definition.

Snapshots of the configuration of the mobile particles show that these
particles tend to form clusters, i.e.  they are not randomly
distributed throughout the system. The spatial correlation between
mobile particles is shown in Fig.~\ref{fig3}, where we compare
(cf. inset) $g_{AmAm}(r)$ and $g_{AA}(r)$, the radial distribution
functions for the mobile particles and for the bulk, respectively. (In
the following, ``bulk'' refers to all of the $A$ particles.)  We find
that at short and intermediate distances ($r\leq 4$) the mobile
particles are more strongly correlated than the bulk.  This is
demonstrated more clearly by computing the ratio
$g_{AmAm}(r)/g_{AA}(r)$, which is shown in Fig.~\ref{fig3} for three
different $T$. From this figure we see that with decreasing $T$ the
relative correlation between the mobile particles increases; i.e., the
relative height of the first nearest neighbor peak ($r\approx 1$)
increases quickly and the ratio decays more slowly as a function of $r$
if $T$ is decreased. At the lowest $T$, the size of the cluster is on
the order of 3 -- 4$ \sigma_{AA}$~\cite{donati97}.  If we assume a
molecule of diameter 0.4 -- 0.5~nm we find that the clusters have a
size of about 1 nm, which is in rough agreement with experimental
expectations \cite{hetero_exp,size-ediger}.  We note that at small
wave-vectors the partial structure factors for the bulk do not show any
indication for the presence of these clusters.  Thus it is perhaps not
surprising that no evidence for the presence of such clusters was found
from the structure factors measured in the neutron scattering
experiments of Leheny, {\it et al.}  \cite{leheny96}.

What is the effect of these clusters of mobile particles on the bulk
relaxation dynamics? To explore this question we compute the
incoherent intermediate scattering function $F_s^{(Am)}(q,t)$ for the
mobile particles and compare it with that for the bulk particles,
$F_s^{(A)}$. These correlation functions, shown in Fig.~\ref{fig4} for
three $T$, are calculated at a wave-vector $q=7.2$, which coincides
with the location of the main peak in the structure
factor~\cite{kob}. From this figure we see that $F_s^{(Am)}$ decays
faster than $F_s^{(A)}$ and that the ratio between the relaxation time
of the two correlation functions increases with decreasing
temperature.  (The relaxation time could, e.g., be defined as the time
it takes a correlation function to decay to $e^{-1}$ of its initial
value.) This ratio is approximately 3 for the highest $T$ and
approximately 10 for the lowest $T$. It is not unreasonable to
extrapolate that at temperatures close to $T_g$ the ratio between the
relaxation times become as large as $10^2-10^4$, similar to values
reported from experiments~\cite{schmidt91,hetero_exp,cicerone97}.
It is also interesting to note that the $\alpha$-relaxation time of 
$F_s^{(Am)}$ is on the order of the end of the $\beta$-relaxation of the 
bulk. This suggests that the relaxation of these clusters might be related to
the $\beta$-relaxation of the bulk.

Also shown in Fig.~\ref{fig4} is the probability $P(t)$ that a particle
which was mobile at time $t=0$ is still mobile at time $t$, at the
lowest $T$ investigated (bold dotted curve). We define $P(t)$ by
$\langle (N(t)-N(0)^2/N_A)/(N(0)-N(0)^2/N_A \rangle$, where $N_A$ is
the total number of $A$ particles and $N(t)$ is the number of particles
that were mobile at time $t=0$ and still mobile at time $t$.  A related
correlation function for the least mobile particles has been measured
in experiments \cite{schmidt91,hetero_exp}.  We see that $P$ decays on
the time scale of the intermediate scattering function of the mobile
particles, demonstrating that the lifetime of a cluster is of the order
of the relaxation time of the particles which constitute the cluster.
However, the lifetime is significantly shorter than the
$\alpha$-relaxation time of the bulk.  More details on the dynamics of
the particles within the clusters will be given
elsewhere~\cite{donati97}.

We find that the existence of clusters of mobile particles is related
to small, local equilibrium fluctuations in
composition~\cite{donati97b}.  At all $T$, the pair correlation
function $g_{AmB}(r)$ between the $B$ particles and the mobile
particles is smaller than the bulk quantity $g_{AB}(r)$ for $r\leq 3$.
Thus mobile particles have fewer $B$ particles in their vicinity than
do generic $A$'s.  Because in this system the attractive interaction
between $A$ and $B$ particles is stronger than either the $AA$ or $BB$
interaction, the presence of a $B$ between two $A$'s lowers the
potential energy, giving rise to an effective attraction between the
$A$'s. $A$ particles in a $B$-rich region can thus be expected to have
a reduced mobility.  $A$ particles in a $B$-poor region, however, will
have a reduced effective attraction between them, resulting in a higher
mobility~\cite{kinetic}.

Via this mechanism we expect that this sort of dynamical heterogeneity
will occur in other fragile glass-forming systems, since local
equilibrium  fluctuations arising in the arrangements (packing) of the 
molecules will always be present~\cite{packing}.  Energetically
favorable packings will reduce the local mobility, while less favorable
packings will enhance the local mobility. Such correlations have been
observed in recent spin glass simulations \cite{pgjc}.

\acknowledgments We gratefully acknowledge very useful discussions with
A. Zippelius and A. Heuer.  Part of this work was supported by the
Deutsche Forschungsgemeinschaft under SFB~262.  PHP acknowledges the
support of NSERC (Canada).

\section*{Figures}
\begin{figure}
\caption{Non-gaussian parameter $\alpha_2$ versus time $t$ for $T=0.550$,
$T=0.480$, and $T=0.451$. The arrows mark the location of the maximum,
i.e. of $t^*$.}
\label{fig1}

\caption{($G_s(r,t)-G_s^{g}(r,t))/G_s^{g}(r,t)$ versus r for
$t=t^*$ for $T=0.550$, $T=0.480$, and $T=0.451$. The arrow marks the
location of $r^*$ for $T=0.451$. Inset: the same quantity on
a logarithmic scale.}
\label{fig2}

\caption{Inset: radial distribution function $g_{AmAm}(r)$ 
and $g_{AA}(r)$ for the mobile and bulk particles,
respectively, for $T=0.451$. Main figure: Ratio between 
$g_{AmAm}(r)$ and $g_{AA}(r)$ for $T=0.550$, $T=0.480$, and $T=0.451$.}
\label{fig3}

\caption{The incoherent intermediate scattering function for the mobile
(bold lines) and the bulk particles (thin lines) for $T=0.550$ (dashed
line), $T=0.480$ (solid line), and $T=0.451$ (dashed-dotted line).
Bold dotted line: Probability $P$ that a particle which is mobile at
time zero is also mobile at time $t$ for $T=0.451$.}
\label{fig4}

\end{figure}

\begin{references}
\bibitem{path} The ($P,T,\rho$) coordinates of the two endpoints of the path are 
(0.489,0.550,1.09) and (2.690,0.451,1.19).
\bibitem{kinetic}
Note that since a fluctuation in the concentration does not affect the
distribution of the velocities of the individual particles, the mean
kinetic energy of the $A$ particles in such a region is the same as the
one in the bulk, thus the enhanced mobility is {\it not} due to a
larger velocity of the mobile particles.
\end{references}
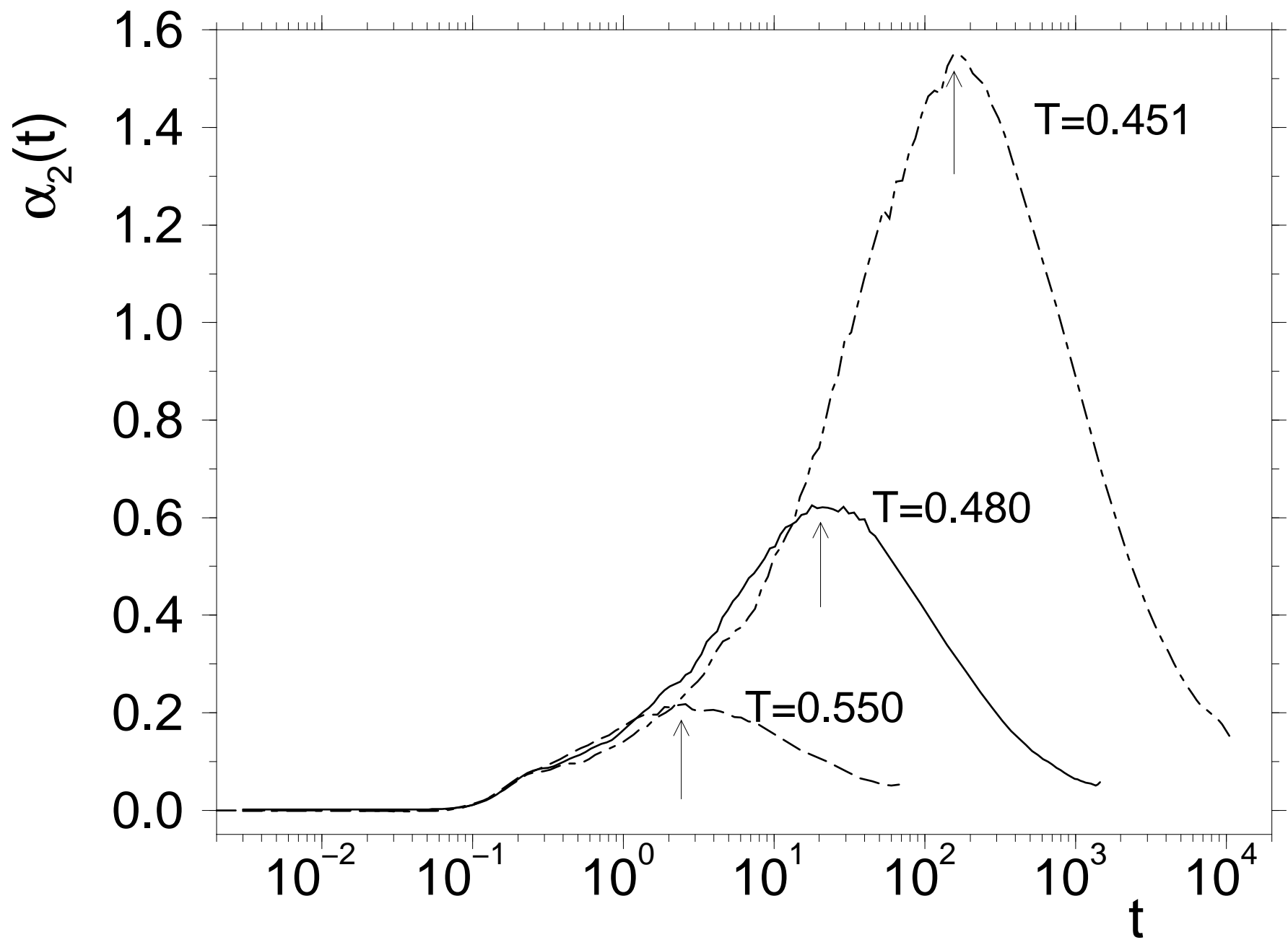
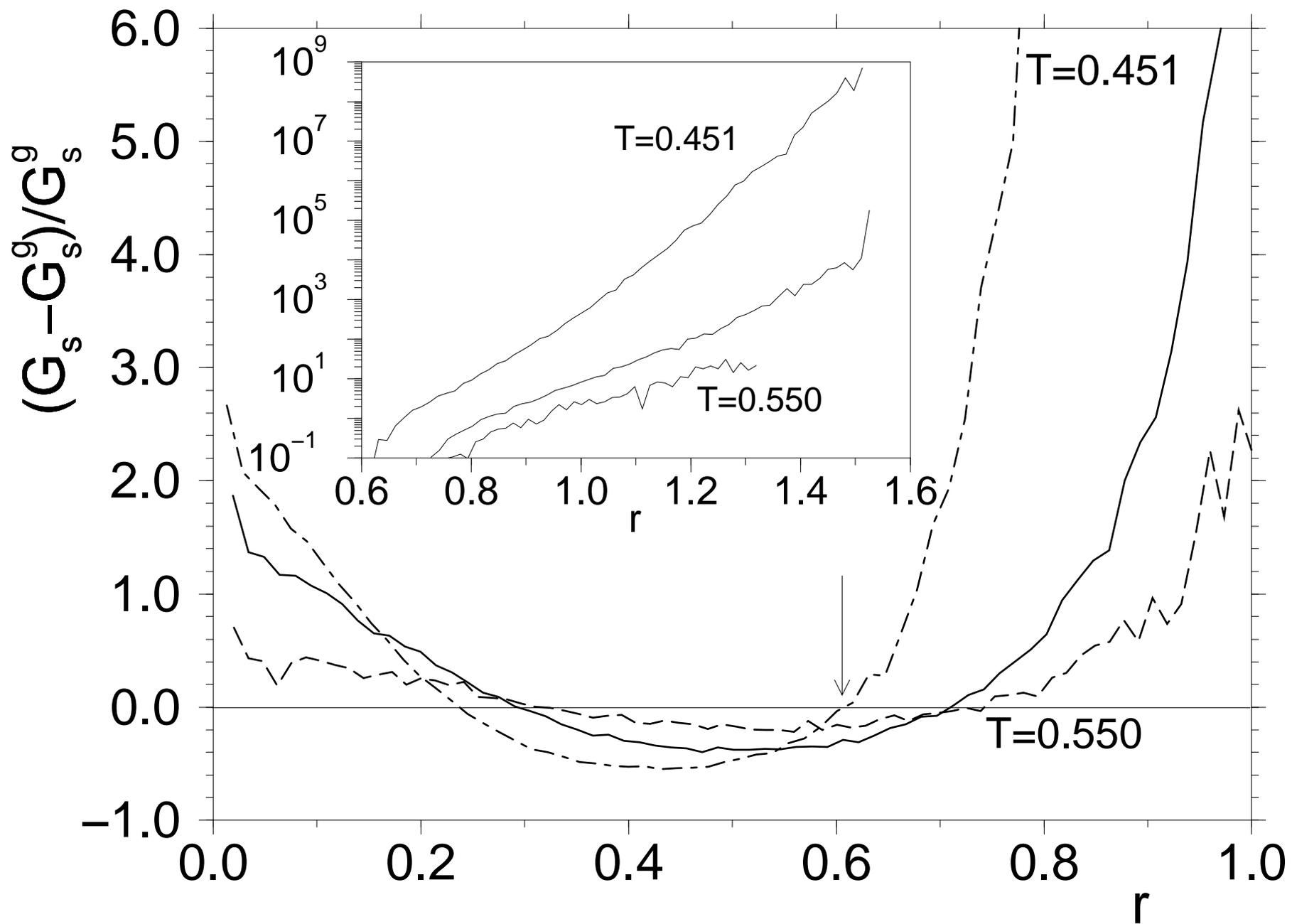
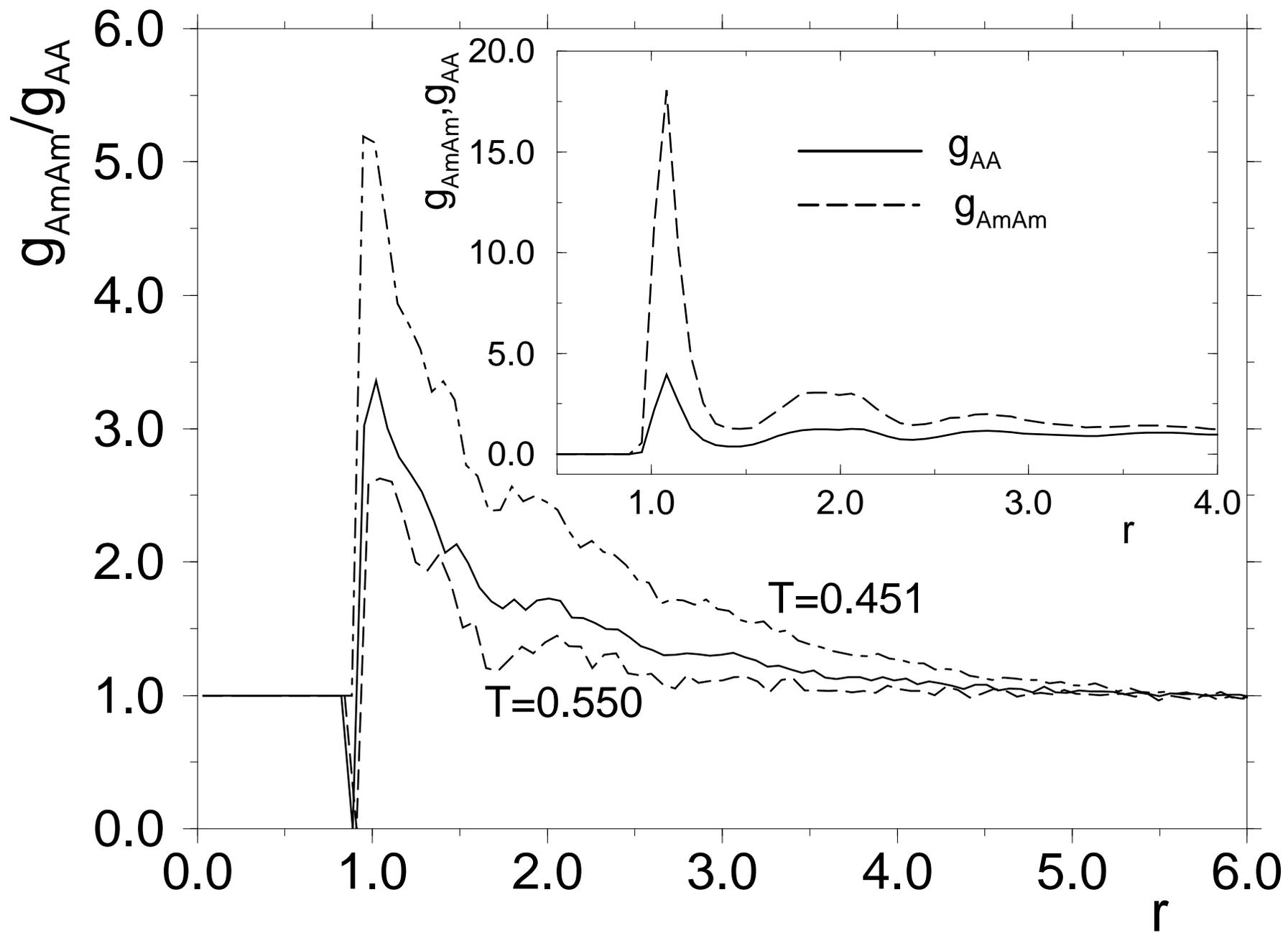
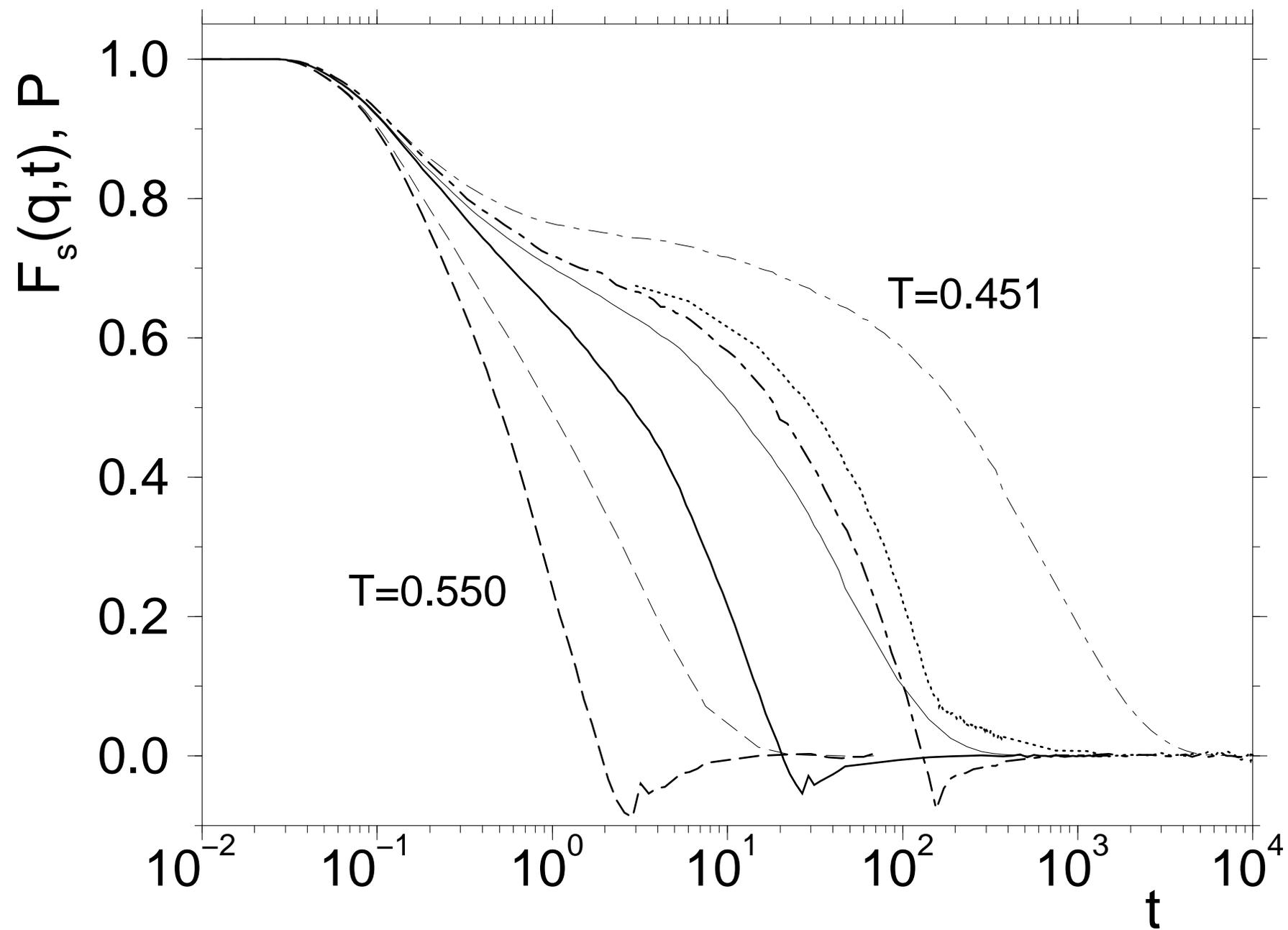
\end{document}